\documentclass[useAMS,usenatbib]{mn2e}
\usepackage{times}
\usepackage{rotating}
\usepackage{subfigure}

\def \kms{\ifmmode{~{\rm km\,s}^{-1}}\else{~km~s$^{-1}$}\fi}
\def \vhel{\ifmmode{V_{{\rm hel}}}\else{$V_{{\rm hel}}$}\fi}
\def \vsys{\ifmmode{V_{{\rm sys}}}\else{$V_{{\rm sys}}$}\fi}
\def \vobs{\ifmmode{V_{{\rm obs}}}\else{$V_{{\rm obs}}$}\fi}
\def \degree{\ifmmode{^{\circ}}\else{$^{\circ}$}\fi}
\def \lsun{\ifmmode{{\rm\ L}_\odot}\else{${\rm\ L}_\odot $}\fi}
\def \msun{\ifmmode{{\rm\ M}_\odot}\else{${\rm\ M}_\odot$}\fi}
\def \myr{\ifmmode{{\rm\ M}_\odot{\rm\ yr}^{-1}}\else{${\rm\ M}_\odot$ 
yr$^{-1}$}\fi}
\def \teff{\ifmmode{{\rm{T}}_{\rm eff}}\else{${\rm{T}}_{\rm eff}$}\fi}
\def \mdot{\ifmmode{{\rm\dot{M}}}\else{${\rm\dot{M}}$}\fi}
\def\mnras{MNRAS}
\def\apj{ApJ}

\def\aap{A\&A}

\def\aj{AJ}

\newcommand{\ha}{H$\alpha$}

\newcommand{\niil}{[N\,{\sc ii}]\ 6584\,\AA}
\newcommand{\niib}{[N\,{\sc ii}]\ 6548\,\AA}
\newcommand{\niiab}{[N\,{\sc ii}]\ 6548,\ 6584\,\AA}

\def \st{\ifmmode{^{\mathrm{st}}}\else{${^{\mathrm{st}}}$}\fi}
\def \nd{\ifmmode{^{\mathrm{nd}}}\else{${^{\mathrm{nd}}}$}\fi}
\def \rd{\ifmmode{^{\mathrm{rd}}}\else{${^{\mathrm{rd}}}$}\fi}
\def \th{\ifmmode{^{\mathrm{th}}}\else{${^{\mathrm{th}}}$}\fi}
\newcommand{\hnii}{{\rm H}$\alpha+$[N~{\sc ii}]}

\title[Halo of Helix nebula]
{The bow-shock and high-speed jet in the faint, 40 arcmin diameter, 
outer  halo of the
evolved Helix planetary nebula (NGC~7293)}
\author[ Meaburn, Boumis \& Akras]{J. Meaburn$^{1}$\thanks{E-mail:
jmeaburn@jb.man.ac.uk}, P. Boumis$^{2}$ \& S. Akras$^{2}$\\
$^{1}$Jodrell Bank Centre for Astrophysics, University of Manchester,
Oxford Rd., Manchester, UK. M13 9PL.\\
$^{2}$Institute of Astronomy, Astrophysics, Space Applications 
and Remote Sensing, National
Observatory of Athens,\\
 I. Metaxa \& V. Pavlou, P. Penteli, GR-15236
Athens, Greece\\
}

\begin{document}

\date{Accepted 2013 August 13. Received 2013 August 9; in original form 2013 June 29}

\pagerange{\pageref{firstpage}--\pageref{lastpage}} \pubyear{2013}

\maketitle

\label{firstpage}

\begin{abstract}

In previous, very deep, optical images of NGC~7293 both a feature that
has the morphology of a bow-shock and one with that of a jet were
discovered in the faint 40\arcmin\ diameter halo of the nebula. 
Spatially resolved
longslit profiles of the \ha\ and \niiab\ nebular emission lines
from both features have now been obtained.

The bow-shaped feature has been found to have 
\ha\ radial velocities close to the systemic heliocentric
radial velocity, $-$27 \kms,
of NGC~7293 and is faint in the \niiab\ emission lines. Furthermore, 
the full width of
these profiles matches the relative motion of NGC 7293 with its 
ambient interstellar medium consequently it is deduced that 
the feature is a real bow-shock caused by the motion of NGC 7293 as it ploughs 
through this medium.The proper motion of the central star
also points towards this halo  feature which substantiates
this interpretation of its origin.

Similarly \niil\ line profiles reveal that the jet-like filament
is indeed a collimated outflow, as suggested by its morphology, 
at around 300 \kms\ with 
turbulent widths of around 50 \kms. It's low
\ha/\niiab\ brightness ratio suggests
collisional ionization as expected in a high-speed jet.

\end{abstract}

\begin{keywords}
ISM: jets and outflows - planetary nebulae: individual (Helix NGC 7293)
\end{keywords}

\section{Introduction}
The evolved Helix (NGC 7293) planetary nebula (PN) is of particular importance
because of its close proximity to the Sun and hence open to observation
over an unprecedented range of spatial scales. 
\cite{har07} measure a distance of only 219 pc to the white dwarf 
(WD 2226-210; 
\citealt{men88}) progenitor star which, with its surface temperature
of 117,000 K  radiatively ionizes the inner structure of
the  envelope ejected in its Asymptotic Giant Branch (AGB)
phase to give the nebula a  bright 
helical appearance
at optical wavelengths. Originally a  dMe late-type 
companion flare star in a central binary system 
\citep{gru01} was thought to be present
but later \citep{su07} this was ruled out in favour 
of the presence of a 35-150 AU diameter debris disk around WD2226-210.

Observations of the morphology and kinematics of many of the structures
up to a diameter of 25 arcmin of NGC~7293 
have been made on a variety of spatial scales
in many wavelength domains and these are summarised in \cite{mea05b},
\cite{mei05}, \cite{mea08}, \cite{mat09}, \cite{mea10}, \cite{ode04} \& 
\cite{ode07}
and references therein up to these
dates. The consensus of opinion is 
that multiple eruptive events along different axes 
have occurred during the evolution
of the central star as it passed from its AGB phase to its present WD state.

For instance \cite{hug86}, Forveille et al. (1986) and \cite{you99} found two
expanding 
molecular tori emitting the CO lines but with their axes orthogonal
to each other: one axis is aligned with the inner bi-polar lobes
that form the bright helical structure  the other
with the fainter lobes (L1 \& L2 in \citealt{mea08}) that project into 
the halo. This complexity is not unexpected when other evolved PNe are 
considered e.g. see \cite{gui13} for a recent example among many.

\begin{figure*}
\centering
\includegraphics[scale=0.45]{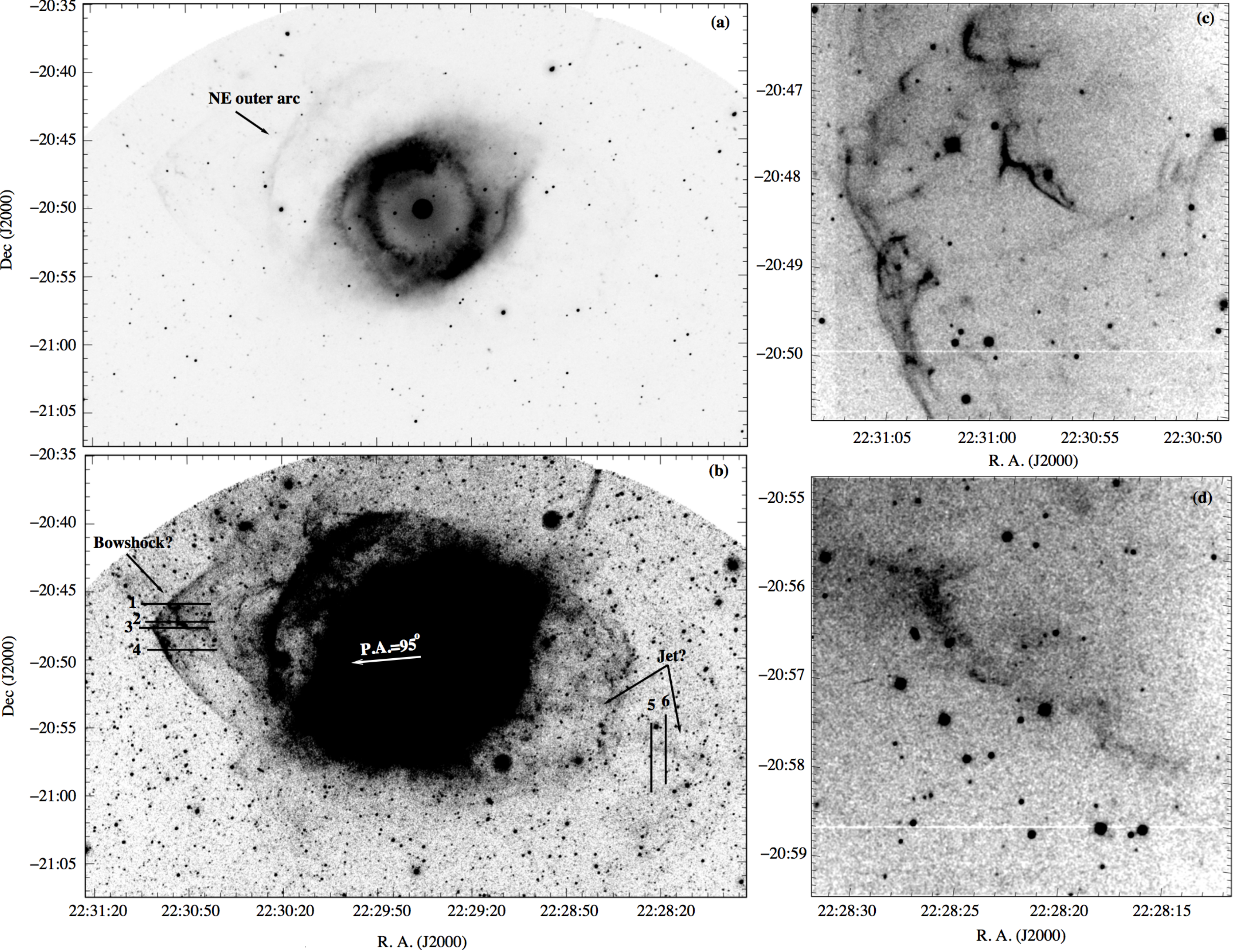}
\caption{a) A light, negative, greyscale presentation of the GALEX NUV
image. The NE outer arc is indicated. b) A deeper representation
of the same image reveals both the NE bow-shaped feature with EW
Slits 1 - 4 marked and the SW jet-like filament with the NS Slits
5 \& 6 marked. The arrowed line over the central star indicates
the direction of it proper motion. c)  \hnii\ image of the bow-shaped region and d) \hnii\ image of the jet-like feature. The horizontal white line in c) and d) is a bad column on the ccd.}
\label{fig1}
\end{figure*}

Jets of collimated ejected material are also not unexpected e.g. see those
 in the PN Fg~1 
\citep{lop93}.
 Remarkably it was a very deep image in the light of the \ha\ plus \niiab\ 
nebular
emission lines obtained with the 30-cm aperture Crete optical telescope
that revealed (see fig. 1c in \citealt{mea05b}) what 
appeared to be a bow-shock and a jet in the very outer 40 arcmin
diameter faint 
halo 
 of NGC 7293. Both of these morphological 
features were detected subsequently 
in the GALEX NUV (175-
280 nm) images and sketched in fig. 6 of \cite{mea08}
where the edge of this halo was called `outer envelope'.
 More recently
the same outer features  have been seen in the WISE all-sky-survey image
of NGC~7293,
at a wavelength of 12 $\mu$m,
by \cite{zha12}.

Deep, spatially and spectrally resolved, longslit profiles of the 
\ha\ and \niiab\  emission lines have now been obtained over both this faint 
 bow-shaped feature  and that with a jet-like appearance in the outer halo
of NGC 7293.
Jets, in particular are expected to manifest themselves as high-speed, 
collimated
outflows and should be easily verified unambiguously by their 
kinematical signature if tilted to the 
plane of the sky.

\subsection{Observations}

The longslit spectral observations were obtained with the second Manchester
Echelle Spectrometer (MES--SPM; \citealt{mea03}) first used
in La Palma but now installed
at the f/7.9 focus of the 2.1-m San Pedro Martir UNAM telescope. 
This spectrometer has no cross-dispersion. For the present observations
a filter of 90 $\AA$ bandwidth was used to isolate the 87$^{th}$ echelle order
containing the nebular \ha\ and \niiab\ emission lines.

A Marconi e2v CCD with 2048 $\times$ 2048 
square pixels, each with 13.5 $\mu$m sides,
was the detector. Two times binning was used in both the spatial and spectral 
dimensions consequently the 1024 increments along the slit length 
are each 0.352 $\arcsec$ long.
A total projected slit length on the sky of 5.12$\arcmin$\ was therefore
limited by the 
30 mm length of the  slit.
`Seeing' varied between 1-2$\arcsec$ during these observations between 
19-26 November 2012.
 The slit was
300 $\mu$m wide ($\equiv$~20 \kms and 3.8~\arcsec) and the integration time
was 1800~s for each slit position. 
The spectra were calibrated to $\pm$~1~\kms\ accuracy
against the Th/Ar arc lamp.

A light, negative, greyscale 
representation of the GALEX NUV image is shown in Fig. 1a to
show the brighter central structures of NGC~7293. This can be 
 compared in Fig. 1b 
with a deeper representation of the same image. The bow--shaped structure
in the NE quadrant and the apparent 
jet-like feature to the SW  can now be seen in Fig. 1b where 
the slit positions 1 -- 6 for the present observations are marked. 
Details of the bow-shaped and jet like
regions can be seen in the images in Fig. 1 c and d. These were obtained with the 2.3-m Aristarchos f/8 telescope at Helmos Observatory on August 6 and 9, 2013, through a 40 \AA\ bandwidth
filter centred on the \hnii\ optical emission lines.  Exposures of 1800 s duration were obtained with a 1024 $\times$ 1024, 24 $\mu$m$^2$ CCD detector ($\equiv$ 0.28 arcsec$^2$) resulting to a 5 $\times$ 5 arcmin$^2$ field of view. The image reduction was carried out using the IRAF and STARLINK packages.

%

\section{Results} 

\begin{figure}
\centering
\includegraphics[scale=0.65]{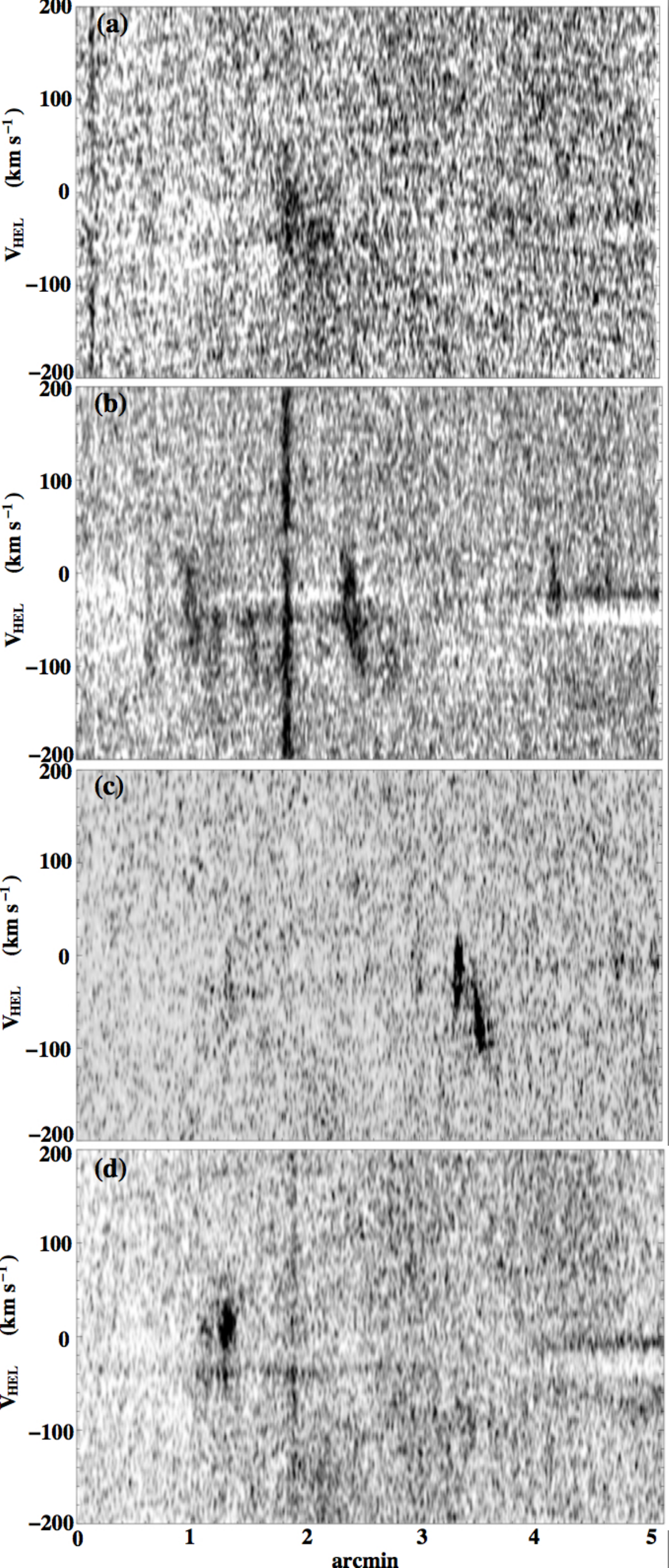}
\caption{a)- d) show negative greyscale representaions of the 
\ha\ line profiles along the
whole slit lengths 1-4 respectively which are marked in 
Fig. 1b. The dark vertical line in b) is the spectrum of a star. 
Faint horizontal features are the residuals of airglow lines
after their partial subtraction from the data arrays.}
\label{fig2}
\end{figure}

\begin{figure*}
\centering
\includegraphics[scale=0.5]{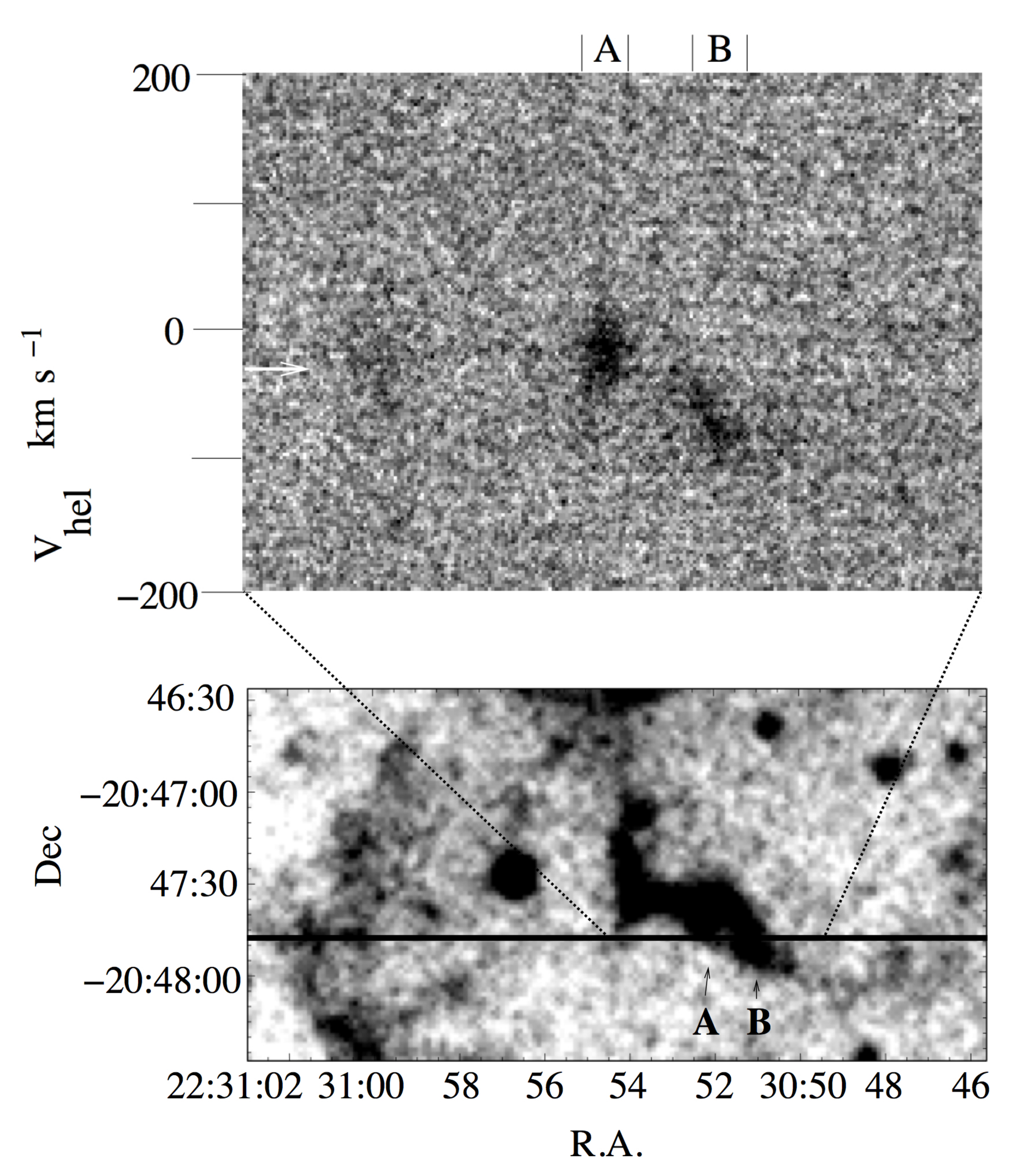}
\caption{A negative greyscale representation of the position-velocity
array of \ha\ line profiles from the EW Slit 3 in Fig. 1b is compared with
an enlargement of part of the GALEX NUV image which contains
the NE bow-shaped region.   
}
\label{fig3}
\end{figure*}

\begin{figure*}
\centering
\includegraphics[scale=0.45]{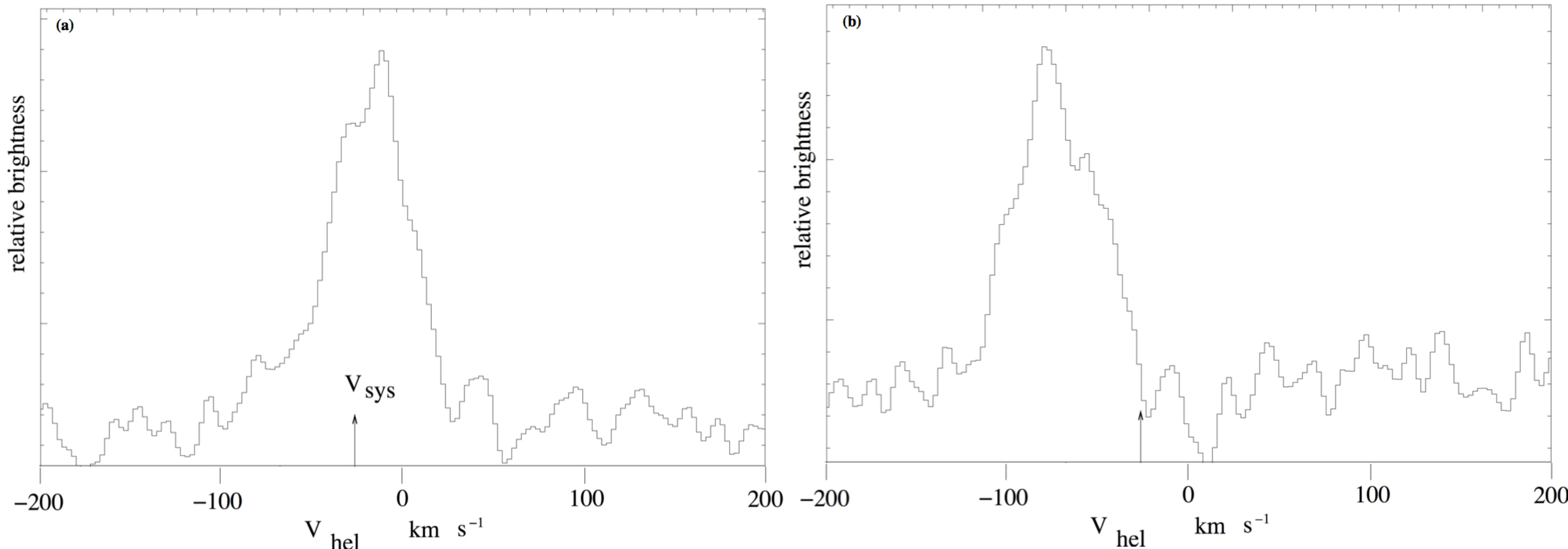}
\caption{The \ha\ line profiles from incremental lengths A \& B
marked in Fig. 3 are shown here in a) and b) respectively. 
}
\label{fig4}
\end{figure*}

\subsection{NE bow-like feature}
Negative grey-scale representations of the 
 position-velocity (PV) arrays of \ha\ profiles 
along the whole lengths of
Slits 1-4, marked in Fig. 1b, are shown in Fig. 2 a-d respectively (and see the optical image in Fig. 1 c).
The PV array of \ha\ profiles
along that part of EW Slit 3, which covers the brightest filament,
is compared in Fig. 3 with an enlargement
of the GALEX image in Fig. 1 for 
the region over NGC~7293
where it was obtained. The \ha\ line profiles in Figs. 4 a \& b are 
for the data in the incremental lengths A and B respectively marked
in Fig. 3. Single Gaussian fits to these 
profiles have turbulent half-widths (fwhm)  of 38~$\pm$~3 \kms\ 
and 49~$\pm$~4~\kms\ 
respectively
when corrected for the 20 \kms\ instrumental width, the 
21.4~\kms\ thermal width of the \ha\ line at 10,000~K and the 
fine structural broadening  of this line of 6.4~\kms.
See \cite{mea91} for details of these corrections 
when applied to the observed \ha\ and \niiab\
line profiles.

The A and B profiles are centred on 
\vhel\ $= -$18~$\pm$ 3 and $-$73~$\pm$~4 \kms\ 
respectively. Their 
central values of \vhel\ should be compared to the systemic heliocentric
radial velocity
 \vsys\ $= -$27~$\pm$~2 \kms\ for
the whole nebula (see fig. 8 in \citealt{mea05b}). 
The many other spectral 
features from the other slit positions over this eastern 
edge (the EW orientated Slits 1, 2 and 4
in Fig.1) can be seen in Fig. 2 a - d to 
have very similar widths and central values of \vhel\ 
as they cross the brighter filamentary edges of this eastern extreme
of the outer halo of NGC~7293.
Incidentally, the  \ha/\niiab\ brightness ratios are $\geq$ 4 for all
of these filaments. Only one minor exception is found. That is for the 
small length of Slit 2 as it crosses the filament $\approx$ 1\arcmin\ 
west of the brightest filament  in Fig. 3.
Here the \niil\ profile  is centred on \vhel\ $= -$67~$\pm$~3~\kms\ 
with a corrected
width of
 21~$\pm$~2~ \kms\ and \ha/\niiab\ brightness ratio $\leq$ 0.2. This somewhat 
anomalous 
spectral feature
is not detected in \ha\ profiles in the 
same PV array but appears faintly in the \niib\ nebular line
profiles.

The \ha\ line profile over the very apex of the bow-shaped feature is
marginally detected in the most easterly parts of Slits 2 and 3 
with about the same radial velocities and spectral 
half-widths of the profiles from incremental lengths A 
and B in Fig. 4a \& b.   

\begin{figure*}
\centering
\includegraphics[scale=0.5]{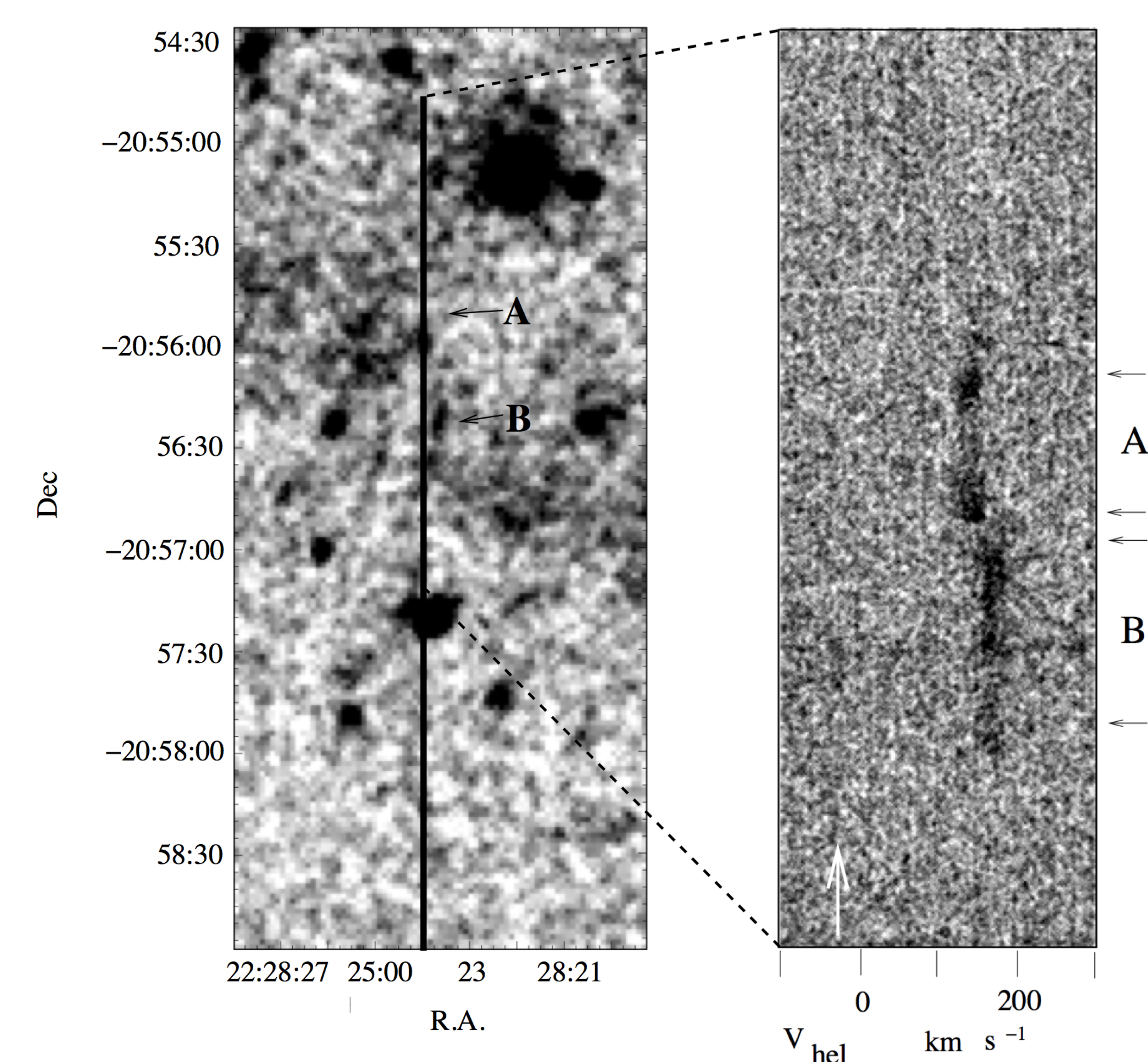}
\caption{A negative presentation of the position-velocity array of \niil\
line profiles from the NS Slit 5 in Fig. 1b is compared
with an enlargement of that part of the GALEX NUV image which contains
the SW jet-like filament. 
}
\label{fig5}
\end{figure*}

\begin{figure*}
\centering
\includegraphics[scale=0.45]{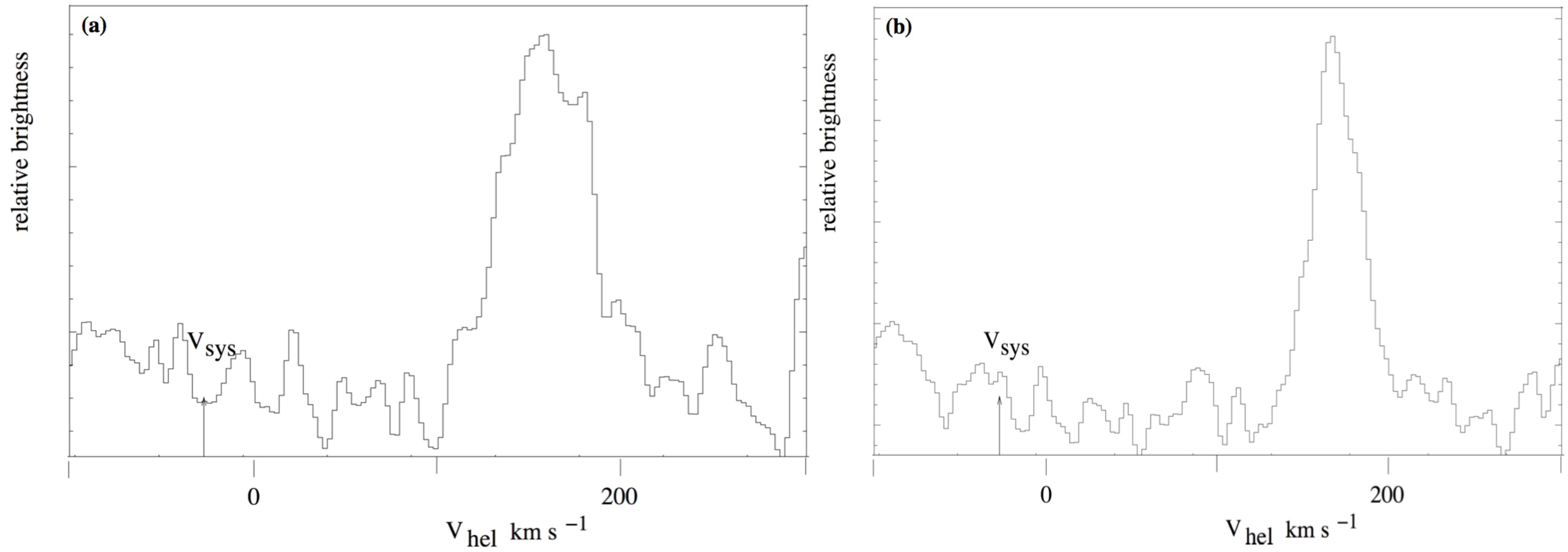}
\caption{The \niil\ line profiles for incremental lengths marked
A \& B in Fig. 5 are shown in a) and b) respectively. 
}
\label{fig6}
\end{figure*}

\subsection{SW jet-like feature}

The negative grey-scale representations of 
PV arrays of \niil\ line profiles from the NS Slit 5 
are compared in Fig. 5 with an enlargement of the GALEX NUV image that 
contains 
the jet--like feature. It can be seen that this feature emits the 
\niil\ line over an 80\arcsec\ length of the NS slit (and see the optical image in Fig. 1 d). Profiles from the
incremental lengths A and B in Fig. 5 are shown in Figs. 6 a \& b 
respectively. Single Gaussian fits to the profiles from A and B indicate
turbulent 
widths (fwhm) of 
53~$\pm$~3 \kms\  (fwhm) and  24~$\pm$~2 \kms\ respectively
 when both widths are corrected for the
instrumental  broadening of 20~\kms\ and the 
\niil\ (10000~K) thermal broadening of 5.7~\kms.
Again see \cite{mea91} for details of these
corrections but note that the \niil\ line, unlike \ha, has no
fine structural components and at a given temperature
is thermally broadened by 14$^{1/2}$ less than \ha.

The profiles from A and B  are centred on 
\vhel\ $=$ 160~$\pm$ 4~\kms\  
and 170~$\pm$ 
3 \kms\ 
respectively.
A similar, but fainter,
spectral feature is seen along Slit 6 centred on \vhel\ $\approx$
 156~\kms. 
The corresponding spectral features in the \niil\ line
are only marginally detected in the
\ha\ line profiles in the same PV arrays 
which indicates a brightness ratio of \ha/\niiab\ $\leq$
0.2. Again these central values of the profiles 
should be compared with \vsys\ $=-$27~\kms\
for the whole NGC~7293 nebula. Collimated jets with velocities of a few hundreds of \kms\ have also been found to other PNe (i.e. M 1-32; \citealt{akr12}).

\section{Discussion}

\subsection{The jet}

Undoubtedly the kinematics described in Sect. 2.2 
of the linear emission line feature, tentatively identified in \cite{mea05b} 
from its morphology alone as a jet in the outer halo 
of NGC~7293, confirms this classification. Its measured 
radial velocity difference 
from \vsys\
of 192 \kms\  converts to an outflow velocity in a direction away from 
the central star of V$_{j} =$ 192/sin($\theta$) \kms\ where $\theta$ is the
angle of the jet to the plane of the sky. The low \ha/\niiab\ brightness
ratio of the jet's line emission also suggests that it is 
composed of collisionally 
ionized, nitrogen-enriched, collimated outflowing 
material originating in the star.

This jet extends to 23.8\arcmin\ from the central star and its tip
is coincident with the  furthest extent of the faint halo of
NGC~7293 (called the 'outer envelope' in \citealt{mea08}). It may
extend beyond this point but not encounter sufficiently dense material
to produce its ionization, consequently a lower limit
to its dynamical age is given by 
T$_{d}$~$\geq$~7400~$\times$~tan ($\theta$) yr for 
a distance to NGC~7293 of 219 pc. 
The value of $\theta$ can only be guessed at but a reasonable value of
$\theta$ $\approx$ 40\degree\ 
gives V$_{j}$ $\approx$ 300 \kms and T$_{d}$~$\geq$~6200 yr. Within all of the
uncertainties of applying this estimation to a jet this is comparable
to T$_{d} =$ 10000 yr estimated by \cite{you99} for the inner
CO emitting torus.

The jet's limited apparent length of $\approx$ 8\arcmin\ starting from 
the `SW outer arc' (see \citealt{mea08}) of the halo of NGC~7293 
suggests that it was a short--lived
event in the evolution of the central star.
It is interesting to speculate about the origin of this truncated
jet within the evolutionary history of NGC~7293. \cite{sok13}
show the presence of bi--polar jets in a pre-planetary nebula
GRL 618 just before the ejection of the asymptotic giant
branch (AGB) wind. Previously, 
\cite{sah98}, in their survey of young PNe
revealed the presence of many jets in objects in the late AGB and/or
early post--AGB evolutionary phases. It is therefore suggested that
the, albeit mono-polar, jet detected in the present work
had its origin in this early stage of the evoluion of NGC~7293
and ceased as the AGB wind declined and the PN was born. 
This explanation would be consistent with the jet's truncated
nature. The tenuous evidence of a counter-jet in Sect. 2.2 
and see \cite{zha12} needs to be investigated
by further observations. 

\subsection{The bow--shock}

As with the SW jet the kinematics of the bow-shaped feature in the NE quadrant
of the outer halo of NGC~7293 can be the key to determine 
its origin. 
Its bow--shaped morphology makes it unlikely to be caused 
by ejected material from NGC~7293 (and see Sect. 3.3)
therefore two further possibilities
should be considered; it could either be a bow-shock ahead of an approaching
counter jet to the receding one in the SW quadrant (Sect. 3.1) or a bow--shock
caused as the whole of the NGC~7293 nebula ploughs through the ambient
interstellar medium (ISM). If an as yet unseen approaching
counter-jet exists, velocities in a bow-shock
around its leading edge of a few hundreds of \kms\ would be expected whereas,
those for that caused by the  motion of NGC~7293 through the ambient
medium would match the relative velocities. In fact, \cite{har87} determined 
that the full-widths of emission lines
from any bow-shocks caused by supersonic motions  
match their shock velocities.
 
The profiles from Slits 1--5 in Figure 1b whose examples are shown
in Figs. 3 and 4a \& b clearly (Sect. 2.1) have extents very much less than the
$\approx$~300 \kms\ required if the bow-shaped feature is generated by a jet
similar to that described in Sect. 3.1 consequently the magnitude of the 
relative
motions of NGC~7293 with respect to its ambient medium must be explored.

Firstly, for the galactic longitude l $=$ 36.2060\degree\ of NGC~7293
the effects of differential galactic 
rotation ($\vhel~\leq$~0.43~\kms) are negligible.
This then leaves the relative motion of NGC~7293 with respect to the ambient 
medium to be determined
by a combination of the tangential 
velocity V$_{t}$ of the central star WD2226-210  
and the measured \citep{mea05b}
\vsys\ $=-$27 \kms\ of the nebula. Using the proper motions of WD2226-210
determined by \cite{cud74} and \cite{ker08} along with the
parallax distance of 219 pc of \cite{har07} a value of  V$_{t} =$ 40
\kms\ along position angle 
PA $=$ 95.7\degree\ (marked on Fig. 1b)  is given.  The velocity of NGC 7293 
with
respect to its ambient medium is therefore V$_{rel} =$ 48 \kms. Moreover, the PA of the PM
of the central star
points nearly directly to the apex of this bow-shaped feature 
(Fig. 1b) and the
radial velocity of the apex is close to \vsys\ of NGC~7293. 

\cite{har87} predicted the shape and extent of  line profiles
from radiative bow shocks as Herbig--Haro objects ploughed into their
ambient medium. This situation is similar, though of course not
identical, to that proposed here. NGC~7293 is moving at
$\approx$~48 \kms\ with respect to the ambient interstellar gas. 
In the production of a bow shock in the latter this should be combined
with the $\approx$~30 \kms\ expansion velocity. A radiative
bow-shock in the ambient gas would therefore be expected 
to be generated by these
combined velocities in the direction of motion determined by the
proper motion of the central star. \cite{har87},
though again for Herbig--Haro objects, showed  that the full width
of the combined line profile from the shocked region is
equal to the relative motion of the `piston' and the ambient
gas. If this idea is reasonably applied to NGC~7293, then
the full width of line profiles should be $\approx$~80 \kms.
It can be seen in the PV arrays of \ha\ profiles 
in Figs~2 a--d \& 3 and the individual profiles from these in
Figs. 4a \& b that their full widths are from \vhel\ $=-$100 to
20 \kms\ around \vsys\ $=-$27 \kms.  
This range is comparable with the relative velocities and it therefore can 
be
stated conclusively, as a consequence of the present kinematic measurements
that a bow-shocked region caused by the passage of NGC~7293
through its local ISM
is present (as suggested by \citealt{zha12} on 
morphological grounds alone).
Similar features in the outer halos of PNe are common (\citealt{mea05a}
for NGC~6583, \citealt{bou09} for HFG1 and \citealt{war07a, war07b}). 

\subsection{Hubble-type expansion}

It is shown in \cite{mea08} that the expansions 
of the inner and outer molecular tori, 
the system of cometary globules, the inner bipolar lobes 
culminating in the north-eastern outer arc (12.5\arcmin\ radius
marked in Fig. 1a) all follow
a Hubble-type law i.e. their expansion velocities increase linearly with 
their distances from the central star with a gradient of $\approx$
7 \kms(arcmin radius)$^{-1}$. This indicates that these are phenomena
ejected from the central star 
over a short period of time with the fastest travelling
furthest.

The jet (Sect. 3.1) with its measured radial velocity of 192 \kms\ 
relative to \vsys\ and extending out to
an apparent radius of 23.8\arcmin\ fails by a factor $\approx$ 2 to fit into
this pattern but this is not too surprising for it has a completely 
different
dynamical origin to that of the ejected shells and tori. 
However, a similar comparison strengthens the suggestion
that
the NE bow-shaped filament is a bow-shock generated by NGC~7293 ploughing
through the local ISM (Sect. 3.2). Its furthest extent from the central
star is 21\arcmin\ and if following the Hubble-law for the inner
ejected material  an expansion velocity of $\approx$~150 \kms\ would be
expected. The largest expansion velocity possible from the present
measurements  (Sect. 2.2) is
$\geq$ 5 times smaller than this. Not only is a counter--jet origin for
the creation of this outer bow-shaped filament ruled out
(Sect. 3.2) but also an ejected origin.

\section{Conclusions}

The jet--like feature in the SW outer quadrant of the giant halo 
of NGC 7293 is confirmed kinematically to have a jet origin for it is
has a collimated outflow velocity of $\approx$~300 \kms.

This jet has an \ha/\niiab\ brightness ratio of $\leq$~0.2 which
confirms its origin as collisionally ionized, nitrogen enriched,
material from the progenitor
star. 

The limited length of the jet's collimated outflow suggests it was
emitted before the inner helical structure of NGC~7293 was ejected.

The bow-shaped outer filamentary structure in the NE quadrant
of the outer halo of NGC~7293 is confirmed kinematically
as a bow-shock created as NGC~7293 ploughs through its ambient
medium. This interpretation is also consistent both 
with the magnitude and direction of the PM of the central star.

\section*{Acknowledgements}
JM is grateful for the hospitality of the National Observatory of
Athens in April 2013 when this paper was initiated. PB would like 
to thank the staff at SPM and Helmos Observatories for their excellent support during 
these observations. The Aristarchos telescope 
operated on Helmos Observatory by the Institute of Astronomy, 
Astrophysics, Space Applications and Remote Sensing of the National 
Observatory of Athens. GALEX (Galaxy Evolution Explorer) is a NASA Small
Explorer, launched in 2003 April. We gratefully acknowledge NASAÕs support 
for construction, operation, and science analysis for the GALEX mission, 
developed in cooperation with the Centre National d'Etudes Spatiales of France
and the Korean Ministry of Science and Technology.

\newpage


%

\end{document}